# Cellular and WiFi Co-design for 5G User Equipment

Yiming Huo, Xiaodai Dong
Department of Electrical and
Computer Engineering
University of Victoria
Victoria, Canada
ymhuo@uvic.ca, xdong@ece.uvic.ca

Wei Xu
National Mobile Communications
Research Laboratory
Southeast University
Nanjing, China
wxu@seu.edu.cn

Marvin Yuen
Viterbi School of Engineering
University of Southern California
Los Angeles, U.S.A
marvinyu@usc.edu

*Abstract*—Motivated by providing solutions to design challenges of coexisting cellular and WiFi for future 5G application scenarios, this paper, first, conducts an in-depth investigation of current technological trends of 5G from user equipment (UE) design perspective, and then presents a cost-effective cellular-WiFi design methodology based on the new distributed phased array MIMO (DPA-MIMO) architecture for practical 5G UE devices as an example. Furthermore, additional 5G cellular-WiFi application scenarios and co-operation details within 5G heterogeneous networks are unveiled on top of the said cellular-WiFi co-enabled 5G UE design.

## I. INTRODUCTION

5G has seen exceedingly rapid growth and promising commercial deployment in recent years. Both academia and industry are accelerating the progress of 5G evolution with enormous efforts. On the aspect of standardization, as early as 2015, three key principle usage scenarios, namely, Enhanced Mobile Broadband (eMBB), Ultra Reliable Low Latency Communications (uRLLC), and Massive Machine Type Communications (mMTC), have been defined by the International Telecommunication Union (ITU) and followed by many organizations and groups [1]. In July 2016, the Federal Communication Committee (FCC) adopted a new Upper Microwave Flexible Use Service [2]. Most recently in December 2017, the first 5G new radio (NR) specifications have been finally approved by the 3GPP [3], which marks a milestone for future large-scale trial experiments and wide commercial deployment.

On top of the standardization progress, as 5G heterogeneous networks become an immediate reality [4], [5], the application and usage scenarios will be largely enriched and thus become more diverse and complicated than ever. In particular, there have been and will be more spectrum enhancement techniques such as, carrier aggregation (CA) and spectrum sharing paradigms represented by LTE licensed assisted access (LTE-LAA) [6]. Implementing as many wireless standards and technologies as possible on one single base station (BS) or user equipment (UE) is ultimately desired, but technically challenging and commercially expensive, considering that both 5G cellular licensed high bands (HBs) such as 28, 37, 39 GHz and WiFi mmWave bands (57-71 GHz) pose very imminent challenges. Furthermore, realizing multi-function, multi-standard user equipment is even more difficult as it is largely constrained by limited hardware resources, slow-growing battery performance, and strict form factor requirements.

In this paper, we initiate an investigation and analysis of contemporary and future wireless user equipment design. Additionally, we unveil detailed circuit and system designs for critical, cost-effective cellular-WiFi reuse with multiplexing techniques and architecture. The remainder of this paper is arranged as follows: Section II thoroughly reviews and analyzes 5G wireless UE design, from both cellular and WiFi aspects; Section III presents brand-new cost-effective cellular/WiFi physical layer design with specific circuits and systems implementation details in a 5G UE; Section IV further presents more details considering the 5G UE cellular-WiFi co-operation within 5G heterogeneous networks, with concluding remarks in Section V.

## II. 5G WIRELESS USER EQUIPMENT DESIGN

5G UE design will be significantly more complicated than current 4G ones in terms of the classic wireless hardware design classfications such as antenna design, radios frequency (RF) design, baseband (modem) design, and PHY-MAC co-design. This evolution is not only a consequence of new 5G technolgogies such as massive MIMO (MaMi) [7], millimeter wave (mmWave) beamforming (BF), and 5G new waveforms, but also the ultimate requirement of ever-growing high-end applications such as wireless virtual reality (VR), ultra-high resolution (UHD) video streaming, vehicle communications, machine learning, etc.

On the aspect of UE cellular design, employing mmWave as 5G high bands brings up seveal major technical challenges including, high propagation loss [8], serious human blockage and human shadowing issues [9], high penetration loss, and weaker diffraction capability due to the stronger particle nature when frequency increases. Consequently, several techniques such as beamforming and MaMi are utilized to deal with these challenges; however these techniques further generate a series of problems for practical hardware designs. Take human blockage for instance, as illustrated in Fig. 1(a), a conventional design method may place the mmWave beamforming module in one specific location, e.g. the central part of the rear housing of 5G smartphone. However, such a design in Fig. 1(a) could lead to a serious human (hand) blockage issue which causes attenuation as large as 30-40 dB [10], [11]. Simply increasing the phased array dimension or effective isotropic radiated power (EIRP) will generate more heat and cause a heat distribution issue. An alternative design as shown in Fig. 1(b) accommodates two BF modules on the top and bottom of the smartphone, which helps partially solve the human hand blockage issue, however the alternative design will be ineffective when the smartphone is horizontally held by two hands. Fig. 1(c) proposes a structure named as a distributed phased array MIMO (DPA-MIMO) [12] where multiple (=8 in Fig. 1(c)) BF modules are arranged in the rear housing. Such design mitigates the human blockage issue, and enhances heat sinking capabilities, while sustaining a faster data rate through enabling higher spatial multiplexing gain. Conventional beamforming design usually employs a direct conversion structure, whereas the DPA-MIMO architecture is composed of multiple mmWave BF modules that realize conversion between 5G high bands and intermediate frequencies (IF).

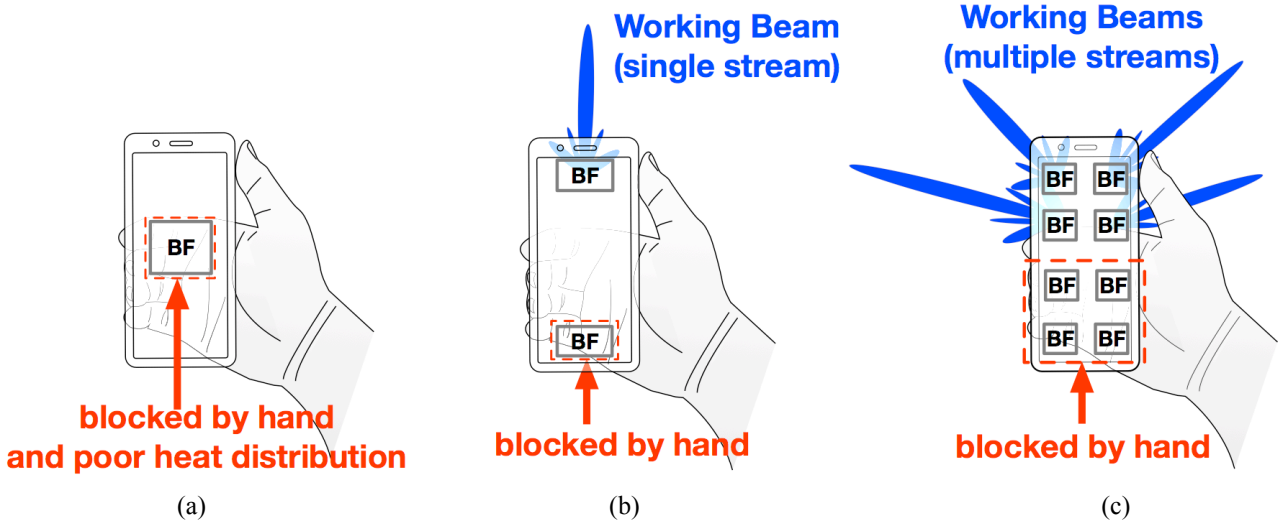

Fig. 1. 5G smartphone with (a) conventional beamforming hardware design, and (b) BF modules on both top and bottom, and (c) DPA-MIMO architecture.

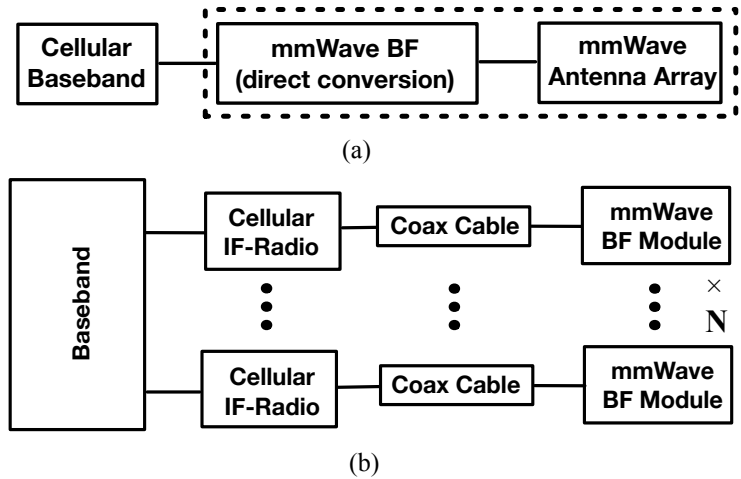

Fig. 2. Wireless hardware block diagram of (a) conventional beamforming design, and (b) DPA-MIMO architecture.

As depicted in Fig. 3, the cellular IF-radio module further process signals between radio frequencies (RF) and a baseband frequency. Coax cables are used to connect BF modules with IF-radio and baseband functional modules, which are accommodated on the main logic board (MLB), and handle transmission precoding [13] and reception combination. Such split-IF architecting, as shown in Fig. 4, can facilitate a highly reconfigurable 5G UE design. As illustrated in Fig. 3, the quantity and placement of BF modules are flexible as long as a necessary edge-to-edge spacing (>1.5 times free-space wavelength [14]) is maintained to guarantee enough spatial isolation and channel capacity. Separating BF and IF modules could also facilitate the reuse of BF and IF modules for other wireless standards.

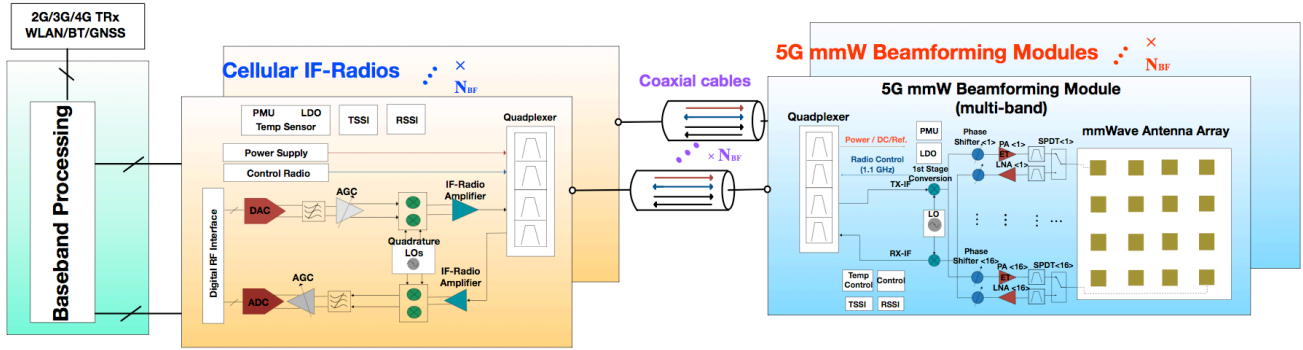

Fig. 3. Circuit and system implementation of DPA-MIMO for 5G UE [12].

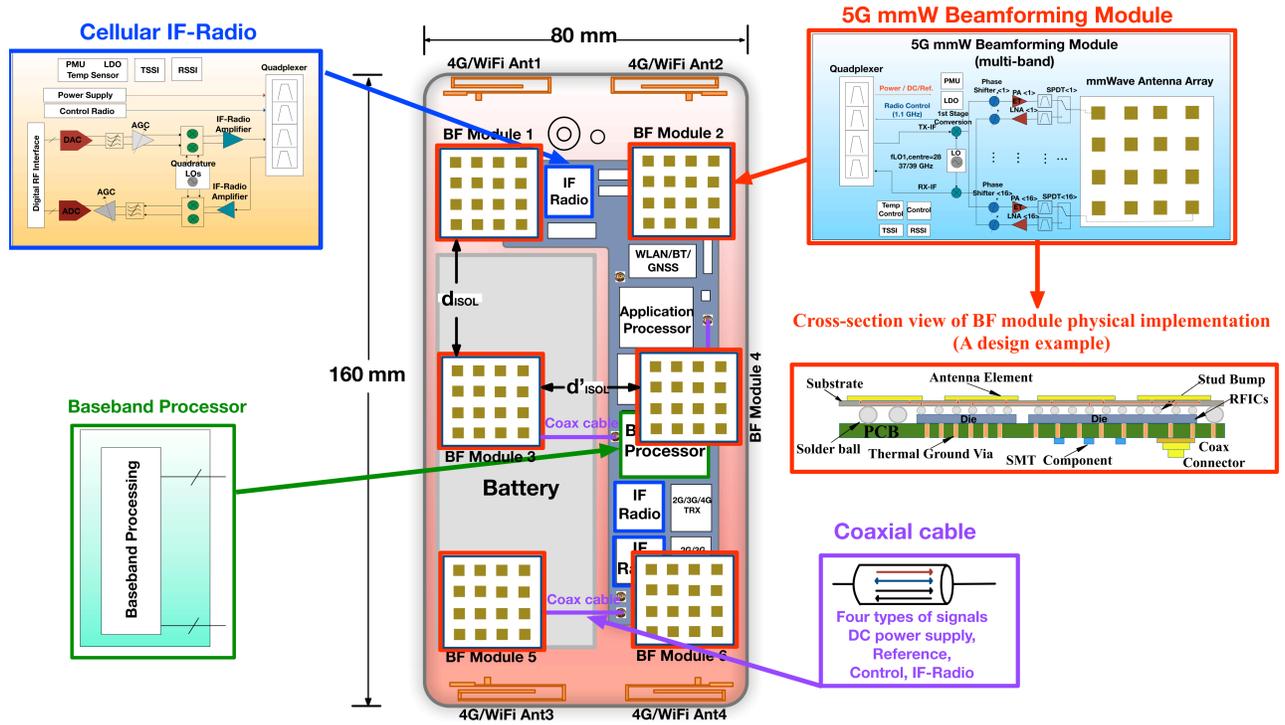

Fig. 4. A 5G UE design example based on DPA-MIMO architecture.

On the other hand, WiFi technology has been evolving significantly and is not only limited to sub-6 GHz bands such as 2.4/5 GHz. With this being said, IEEE 802.11ad/ay standards, known as WiGig, employ 60 GHz bands that are constantly being expanded (57-71 GHz in U.S.), and will play a critical role in future 5G heterogeneous networks. It is also important to point out that, as predicted based on the current LTE-Licensed Assisted Access (LTE-LAA) techniques and standards, 5G-LAA will become an immediate reality (not yet standardized). When 5G cellular mmWave meets WiFi WiGig, they will combine to form more powerful aggregated bands which will realize, at least, a 10 times performance boost compared to the current, most advanced LTE-LAA in terms of the data rate and latency. Therefore, it would be wise to co-exist mmWave cellular, mmWave WiFi, and 5G-LAA altogether, to handle various application and usage scenarios.

III. 5G CELLULAR WIFI CO-DESIGN

Integrating aforementioned 5G wireless technologies and functions on 5G UE devices is a commercial necessity, but costly and technically challenging, particularly considering the limited hardware space at the UE end. Hardware resource competition between 3GPP cellular standards, and IEEE WiFi standards on a smartphone may additionally become a severe problem. For example, when both standards require multiple BF modules, 5G UE designers need to arrange them within a limited hardware area. To further exacerbate the technical challenges, the mmWave BF modules are technically demanding, expensive, and power-hungry.

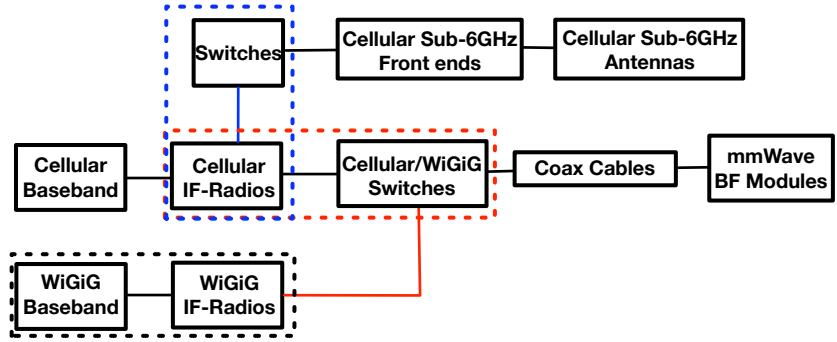

Fig. 5. 5G cellular and WiFi (WiGig) reuse/multiplexing function.

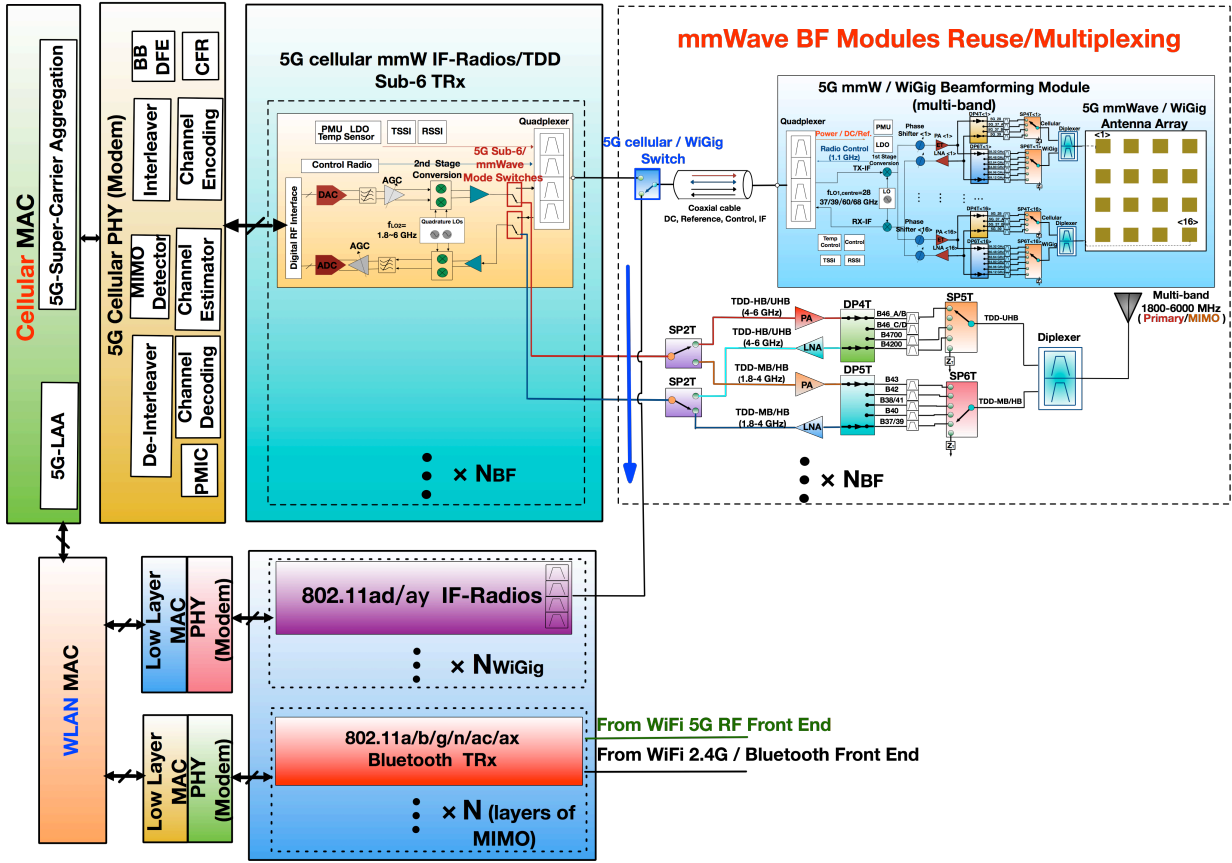

Fig. 6. 5G cellular and WiFi (WiGig) reuse/multiplexing function based on DPA-MIMO.

As illustrated in Fig. 5, a cellular/WiGig mode switch is inserted between coax cables and cellular IF-radios, as well as WiGig IF-radios. As a result, the mmWave BF module can be multipexed for either WiFi WiGig or 5G cellular functionality. Furthermore, cellular IF-radios can be also reused for cellular sub-6GHz front ends and antennas through enabling the switches connected to them. Consequently, 5G mmWave cellular, WiGig, and 5G sub-6GHz cellular can be reconfigured. In particular, WiGig and sub-6GHz cellular functions could be simultaneously activated on request.

A detailed circuit and system design example based on DPA-MIMO architecture is given in Fig. 6. There are a total of $N_{BF}$ groups of 5G mmWave BF modules, cellular IF-radios, and $N_{WiGig}$ WiGig IF-radios. As noted, $N_{WiGig}$ should be no bigger than $N_{BF}$. WiGig and 2.4/5GHz WiFi have different modems and low-layer MAC designs. The cellular modem will be complex as it needs to provide backward compatibilty to 3GPP legacy standards, as well as also support 5G NR that may employ several canidate waveforms [15], such as orthogonal frequency division multiplexing (OFDM) based multicarrier waveforms. Examples of OFDM based multicarrier waveforms include cyclic prefix OFDM (CP-OFDM), Discrete Fourier Transform-spread-OFDM (DFT-s-.OFDM), universal filtered multicarrier (UFMC), filter bank multicarrier (FBMC), generalized frequency division multiplexing (GFDM), and single carrier (SC) waveform such as single carrier frequency division multiple access (SC-FDMA).

On the other hand, it is worthy to mention that, wideband multi-band phased antenna array design [16], [17] and multi-band power amplifier (PA) design [18] are also the critical enabling factors for cellular-WiFi multipexed arthictectures.

IV. 5G CELLULAR WIFI CO-OPERATION

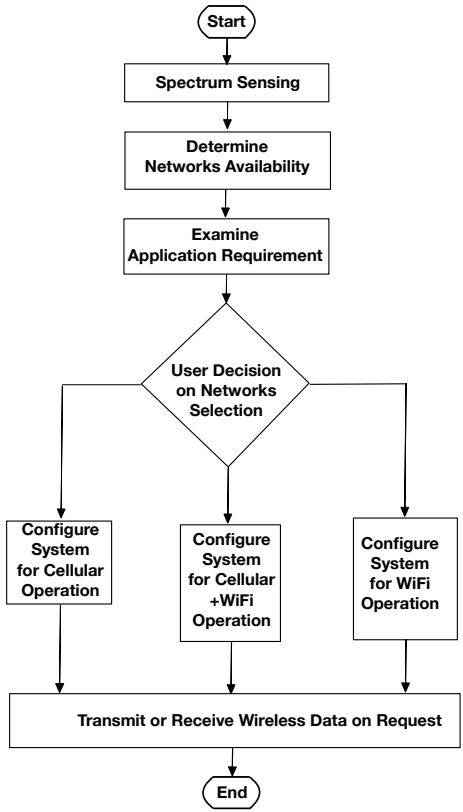

Fig. 7. Flow chart of multiplexed DPA-MIMO wireless communication in a user equipment device.

As concluded from above, a multiplexed DPA-MIMO architecture accomodates a plurality of BF modules, 5G sub-6GHz front ends and antennas, cellular IF-radios, WiFi/WiGig IF-radios, etc. Therefore it can facilitate very rich and diverse application scenarios. A higher MAC layer design and physical-layer-MAC-layer (PHY-MAC) cross-layer design should be carefully considered to enable more cost-efficient co-operation of cellular and WiFi or other wireless technologies. As shown in Fig. 6, a cellular MAC block needs to work with a WLAN MAC block to enable standalone (cellular/WiFi) functions or functions requiring co-operation between cellular and WiFi such as 5G-LAA. Additionally, 5G super carrier aggregation (5G-Super CA), that involves an even wider range of bands, such as from sub-6GHz licensed/unlicensed to above-6GHz licensed/unlicensed, may also be enabled.

Fig. 7 illustrates a flow chart illustrating an exemplary process for multiplexed DPA-MIMO wireless communication which comprises of several steps, namely, spectrum sensing step, determining network availability step, examining application requirement step, network selection step, configuring cellular operation step, configuring cellular and WiFi (5G-LAA, or 5G-Super CA) operation step, and configuring WiFi operation step. More complex and specific algorithms could be built and expanded on this flow chart to enable more cost-effective PHY-MAC cross-layer architectures.

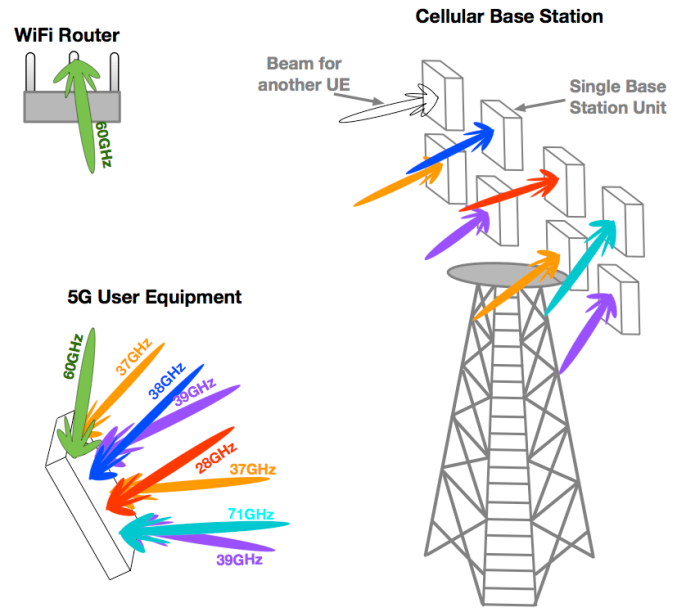

Fig. 8. Wireless communication for a multiplexed DPA-MIMO based 5G UE device within 5G heterogeneous networks.

Furthermore, Fig. 8 depicts one of the situations when wireless communication is established between a multiplexed DPA-MIMO based 5G UE device, a WiFi router, and a cellular base station. In one specific operation mode, multiple BF modules on the UE are enabled to communicate with the cellular base station using different carrier frequencies (particularly for neighboring modules) to mitigate potential interference. Moreover, a 5G-Super CA mode can be operational when unlicensed 71 GHz band and WiFi 60 GHz band are aggregated.

V. CONCLUSION

This paper introduces a series of cellular-WiFi co-design and co-enabling techniques, on top of the cellular-WiFi functional reuse with multiplexing hardware architecture and PHY-MAC cross-layer designs. Furthermore, detailed circuits, system designs, and implementations are specified with detailed cellular-WiFi cooperation details within 5G heterogeneous networks. With the said techniques and architectures, very rich and diverse 5G application and usage scenarios can be facilitated, which address future 5G design and application challenges.

REFERENCES

[1] Rec. ITU-R M.2083-0, "IMT Vision - Framework and overall objectives of the future development of IMT for 2020 and beyond," Sep, 2015.
[2] "REPORT AND ORDER AND FURTHER NOTICE OF PROPOSED RULEMAKING," [Online]. Available: https://apps.fcc.gov/edocs_public /attachmatch/FCC-16-89A1.pdf

[3] "First 5G NR Specs Approved," [Online]. Available: http://www.3gpp.org/news-events/3gpp-news/1929-nsa_nr_5g

[4] C.-X. Wang, F. Haider, X. Gao, X.-H. You, Y. Yang, D. Yuan, H. M. Aggoune, H. Haas, S. Fletcher, and E. Hepsaydir, "Cellular architecture and key technologies for 5G wireless communication networks," *IEEE Commun. Mag.*, vol. 52, no. 2, pp. 122-130, Feb. 2014.

[5] F. Boccardi *et al.*, "Five Disruptive Technology Directions for 5G," *IEEE Commun. Mag.*, vol. 52, no. 2, pp. 74-80, Feb. 2014.

[6] S. Han, Y.-C. Liang, Q. Chen, B.-H. Soong, "Licensed-assisted access for LTE in unlicensed spectrum: A MAC protocol design," *IEEE J. Sel. Areas Commun.*, vol. 34, no. 10, pp. 2550-2561, Oct. 2016.

[7] S. Malkowsky *et al.*, "The World's first real-time testbed for massive MIMO: Design, implementation, and validation," *IEEE Access*, vol. 5, pp. 9073-9088, 2017.

[8] T. S. Rappaport, Y. Xing, G. R. MacCartney, Jr., A. F. Molisch, E. Mellios, and J. Zhang, "Overview of millimeter wave communications for fifth-generation (5G) wireless networks-with a focus on propagation models," *IEEE Trans. Antennas Propag.*, vol. 65, no. 12, pt. I, pp. 6213–6230, Dec. 2017.

[9] G. R. MacCartney Jr., T. S. Rappaport, and S. Rangan, "Rapid Fading Due to Human Blockage in Pedestrian Crowds at 5G Millimeter-Wave Frequencies," in *Proc. IEEE Global Communications Conference (GLOBECOM)*, Dec. 2017, pp. 1-6.

[10] M. N. Kulkarni, A. O. Kaya, D. Calin, and J. G. Andrews "Impact of Humans on the Design and Performance of Millimeter Wave Cellular Networks in Stadiums," in *Proc. IEEE Wireless Communications and Networking Conference (WCNC)*, Dec. 2017, pp. 1-6.

[11] G. R. MacCartney Jr., S. Deng, S. Sun, and T. S. Rappaport, "Millimeter-wave human blockage at 73 GHz with a simple double knife-edge diffraction model and extension for directional antennas," in *Proc. IEEE 84th Veh. Technol. Conf. (VTC-Fall)*, Sep. 2016, pp. 1–6.

[12] Y. Huo, X. Dong, and W. Xu, "5G Cellular User Equipment: From Theory to Practical Hardware Design," *IEEE Access*, vol. 5, pp. 13992-14010, 2017.

[13] L. Liang, W. Xu, and X. Dong, "Low-complexity hybrid precoding in massive multiuser MIMO systems," *IEEE Wireless Commun. Lett.,* vol. 3, no. 6, pp. 653-656, Dec. 2014.

[14] F. Rusek *et al.*, "Scaling up MIMO: Opportunities and challenges with very large arrays," *IEEE Signal Process. Mag.*, vol. 30, no. 1, pp. 40-60, Jan. 2013.

[15] R. Gerzaguet *et al.*, "The 5G candidate waveform race: A comparison of complexity and performance," *EURASIP J. Wireless Commun. Netw.*, vol. 1, pp. 13, Jan. 2017.

[16] C.-X. Mao, S. Gao, and Y. Wang, Q. Luo, and Q.-X. Chu, "Broadband High-Gain Beam-Scanning Antenna Array for Millimeter-Wave Applications," *IEEE Trans. Antennas Propag.*, vol. 65, no. 9, pp. 4864-4868, Sep. 2017.

[17] W. Hong, "Solving the 5G Mobile Antenna Puzzle: Assessing Future Directions for the 5G Mobile Antenna Paradigm Shift," *IEEE Microwave Mag.*, vol. 18, no. 7, pp. 86-102, 2017.

[18] S. Hu, F. Wang, and H. Wang, "A 28GHz/37GHz/39GHz multiband linear Doherty power amplifier for 5G massive MIMO applications," *IEEE Int. Solid-State Circuits Conf. (ISSCC) Dig. Tech. Paper*, pp. 32-33, Feb. 2017.